\documentclass[a4paper,11pt]{article}
\usepackage[utf8]{inputenc}
\usepackage[T1]{fontenc}
\usepackage[english]{babel}
\usepackage[width=160mm,height=247mm]{geometry}
\usepackage{amsmath, amsfonts, amssymb, bm}
\usepackage[pdftex]{graphicx}
\usepackage{lmodern, indentfirst, cite}
\usepackage{siunitx}
\usepackage[textfont={footnotesize},labelfont={color=blue,bf,sf},labelsep=endash]{caption}
\usepackage[colorlinks=true,linkcolor=blue,citecolor=blue,urlcolor=blue]{hyperref}

\usepackage[caption = false]{subfig}
\usepackage{caption}
\usepackage{float}
\usepackage{subfig}
\usepackage{bm}       
\usepackage{cite}     
\usepackage{hyperref} 
\usepackage{braket}
\usepackage[affil-sl]{authblk}
\DeclareMathOperator{\sech}{sech}
\DeclareMathOperator{\csch}{csch}
\usepackage{hyperref}

\title{\textbf{Interplay Between Quantum Coherence and Multiparameter Quantum Estimation in Graphene}}
\author[a]{
Younes Moqine
}
\author[a]{Brahim Adnane
}
\author[b]{	Abdelilah El Rhazali
}
\author[b,c]{Rachid Houça\thanks{r.houca@uiz.ac.ma}
}
\affil[a]{LPMC. Laboratory, Theoretical Physics Group, Faculty of Sciences, Chouaïb Doukkali University, PO Box 20, 24000 El Jadida, Morocco}
\affil[c]{LPTHE. Laboratory, Theoretical Physics and High Energy, Faculty of Sciences, Ibn Zohr University, PO Box 8106, Agadir, Morocco}
\date{}

\newcommand{\beq}{\begin{equation}}
	\newcommand{\eeq}{\end{equation}}

\begin{document}
\begin{titlepage}
	\newgeometry{width=175mm, height=247mm}
    \maketitle
    \thispagestyle{empty}
\vspace{1cm}

\begin{abstract}
In this work, we investigate the relationship between quantum coherence and multiparameter quantum estimation in a graphene-based system. We focus on the estimation of two relevant physical parameters, namely the temperature $T$ and the wave vector $k_x$, and analyze how their variations affect both quantum coherence and the achievable metrological precision. The minimum variances associated with the estimation process are evaluated through the quantum Cramér--Rao bound within both simultaneous and independent estimation schemes. Our results show that quantum coherence is enhanced in the low-temperature regime and around $k_x=0$, while it decreases progressively as either the temperature or the wave vector increases. However, the regions where coherence is maximal do not necessarily coincide with those of optimal estimation precision. In particular, the variance associated with temperature estimation exhibits a divergent behavior near $T=0$, indicating that the system becomes weakly sensitive to small temperature variations in this regime. By contrast, the estimation of the wave vector $k_x$ is more directly related to the coherence properties of the system, with improved precision obtained near $k_x=0$. Furthermore, we introduce the ratio $\Gamma$ to compare the total variances obtained from the independent and simultaneous estimation schemes. This quantity provides a useful measure of the relative difference between the two strategies when the parameters are estimated separately or jointly. The behavior of $\Gamma$ shows that this difference becomes more pronounced for increasing temperature and larger wave-vector values, reflecting the nontrivial role of the multiparameter nature of the estimation problem. These findings demonstrate that quantum coherence can support metrological performance in specific regimes, but it is not sufficient on its own to guarantee optimal precision in multiparameter estimation.

\end{abstract}

\vspace{1cm}

\end{titlepage}

\section{Introduction}

Graphene has attracted considerable attention in condensed matter physics and quantum technologies due to its remarkable electronic properties and its low-energy excitations that behave as massless Dirac fermions~\cite{novoselov2004,geim2007,castro2009}. Since the experimental isolation of graphene and the observation of its field-effect behavior~\cite{novoselov2004}, this two-dimensional material has become a privileged platform for studying relativistic-like quasiparticles in condensed matter systems. The linear dispersion relation near the Dirac points, together with the two-dimensional nature of graphene, makes it highly relevant for exploring fundamental quantum effects and for developing high-sensitivity quantum devices~\cite{geim2007,castro2009}. In particular, the dependence of its quantum state on physical parameters such as temperature and wave vector provides a natural framework for quantum sensing and quantum metrology.

Quantum metrology aims to improve the precision of parameter estimation by exploiting quantum features of physical systems~\cite{helstrom1976,paris2009}. Among these features, quantum coherence plays a central role, since it quantifies the ability of a quantum state to exist in a superposition of basis states. Coherence is now widely recognized as a valuable resource in quantum information theory, thermodynamics, and metrology~\cite{baumgratz2014}. However, the presence of strong coherence does not always guarantee optimal estimation precision, especially when several parameters are estimated simultaneously. In such cases, correlations between parameters and the possible incompatibility of optimal measurements may strongly affect the attainable precision~\cite{ragy2016,liu2020}. Understanding how quantum coherence affects metrological performance in graphene is therefore important for exploring the potential of two-dimensional materials in quantum sensing. In this work, we investigate whether the regions of maximum quantum coherence also correspond to the regions of optimal precision in the estimation of the temperature $T$ and the wave-vector component $k_x$.

The fundamental limits of estimation precision are commonly described by the quantum Fisher information and the quantum Cramér--Rao bound~\cite{helstrom1976,paris2009}. For single-parameter estimation, the quantum Fisher information determines the minimum variance achievable by any unbiased estimator. In multiparameter estimation, this concept is generalized through the quantum Fisher information matrix, whose inverse provides a lower bound on the covariance matrix of the estimated parameters~\cite{liu2020}. Nevertheless, unlike the single-parameter case, the multiparameter quantum Cramér--Rao bound is not always saturable, due to the possible noncommutativity of the symmetric logarithmic derivatives associated with different parameters~\cite{ragy2016,liu2020}.

To address this question, we consider a graphene-based system described by an effective Dirac-like Hamiltonian and analyze its thermal state in terms of the temperature $T$ and the wave-vector component $k_x$. The main objective is to determine how these two parameters influence both the coherence properties of the system and the achievable estimation precision. To this end, we evaluate the quantum Fisher information matrix and the corresponding quantum Cramér--Rao bounds for both simultaneous and independent estimation schemes.

The results reveal that quantum coherence can support metrological performance in specific parameter regimes, particularly for the estimation of the wave-vector component $k_x$. However, maximum coherence does not necessarily coincide with optimal estimation precision, especially for temperature estimation. This highlights the nontrivial interplay between coherence, parameter sensitivity, and multiparameter correlations in graphene-based quantum metrology.

The paper is organized as follows. Section~\ref{sec:framework} presents the theoretical framework for quantum coherence and multiparameter quantum estimation. Section~\ref{sec:model} introduces the graphene model and the corresponding thermal density matrix. Sections~\ref{sec:simultaneous} and~\ref{sec:independent} analyze the precision bounds in the simultaneous and independent estimation schemes, respectively. Section~\ref{sec:ratio} compares the two strategies through a metrological performance ratio. Finally, Section~\ref{sec:conclusion} summarizes the main results and discusses possible perspectives.
\section{Theoretical Framework for Quantum Coherence and Multiparameter Precision}\label{sec:framework}
Quantum coherence and quantum correlations represent fundamental resources in quantum information science. In quantum metrology, these resources may significantly improve the sensitivity of a probe system to small variations of physical parameters. In the present section, we introduce the coherence quantifiers and the quantum Fisher information matrix formalism adopted throughout this work.
\subsection{Quantum Coherence}
Quantum coherence describes the presence of superposition between different states of a chosen reference basis. For a density matrix $\rho$, the amount of coherence is therefore not an absolute quantity, but depends on the basis in which the state is represented. In other words, a state may exhibit coherence in one basis and become diagonal, and hence incoherent, in another basis. This basis dependence is particularly important in physical systems such as graphene, where the choice of basis is related to the effective degrees of freedom used to describe the Hamiltonian and the corresponding thermal state.

Several measures can be used to quantify this property. A standard and widely used choice is the $\ell_1$-norm of coherence, defined as
\begin{equation}
	C_{\ell_1}(\rho) = \sum_{i \neq j} |\rho_{ij}|.
\end{equation}
This expression evaluates the total contribution of the off-diagonal elements of the density matrix in the selected reference basis. Since these elements are directly associated with quantum superpositions between basis states, the $\ell_1$-norm provides a simple and convenient indicator of the coherence contained in the system.

However, it is important to emphasize that the $\ell_1$-norm is mainly useful because of its direct computability and its sensitivity to the magnitude of the off-diagonal terms. Its physical interpretation should therefore be understood in this restricted sense: it quantifies the total weight of the coherences present in the chosen basis, rather than providing a basis-independent measure of quantumness. In the present work, this quantity is used to analyze how temperature and the wave-vector component affect the off-diagonal structure of the thermal density matrix.
\subsection{Multiparameter Precision}
We now present the mathematical background needed to construct the quantum Fisher information matrix. In particular, we employ a vectorization procedure that maps a matrix into a column vector. This technique is useful because it allows the elements of the QFIM to be evaluated without performing an explicit diagonalization of the density matrix. Let $M^{n\times n}$ be the space of $n \times n$ real or complex matrices. For any matrix $A \in M^{n\times n}$, the vec-operator is defined as \cite{gilchrist2009}
\begin{equation}
	\mathrm{vec}[A] =
	(a_{11},\ldots,a_{n1},a_{12},\ldots,a_{n2},\ldots,a_{1n},\ldots,a_{nn})^{T}.
\end{equation}
Furthermore, using the expansion
\[
A=\sum_{k,l=1}^{n} a_{kl} |k\rangle \langle l|,
\]
the vec-operator can also be written in the following form
\begin{equation}
	\mathrm{vec}[A] =
	\left(I_{n\times n}\otimes A\right)
	\sum_{i=1}^{n} e_i \otimes e_i .
\end{equation}
Here, $e_i$ denotes the elements of the computational basis of $M^{n\times n}$. This operation transforms the matrix $A$ into a single column vector by arranging its columns successively. By using the properties of the Kronecker product \cite{schacke2004}, one obtains
\begin{equation} \label{6}
	\mathrm{vec}[AB]
	=
	(I_n \otimes A)\mathrm{vec}[B]
	=
	(B^{T} \otimes I_n)\mathrm{vec}[A],
\end{equation}
\begin{equation} \label{7}
	\mathrm{tr}(A^{\dagger}B)
	=
	\mathrm{vec}[A]^{\dagger}\mathrm{vec}[B],
\end{equation}

\begin{equation} \label{8}
	\mathrm{vec}[AXB]
	=
	(B^{T} \otimes A)\mathrm{vec}[X].
\end{equation}
These identities are valid for arbitrary matrices $A$, $B$, and $X$. Before giving the explicit expression of the QFIM based on the vectorized density matrix $\rho$, let us recall the main concept of multiparameter quantum estimation. We consider a set of unknown parameters $\{\theta_i\}=\{\theta_1,\theta_2,\ldots,\theta_n\}$. The quantum Fisher information quantifies the maximum amount of information that can be extracted about a parameter through an optimal quantum measurement.

For a quantum state $\rho_\theta$ depending on a single parameter $\theta$, the quantum Fisher information is given by
\begin{equation}
	\mathcal{F}(\rho_\theta)=\mathrm{Tr}\{\rho_\theta L_\theta^2\},
\end{equation}
where $L_\theta$ is the symmetric logarithmic derivative. When several parameters $\theta_i$ are involved, the relevant quantity becomes the quantum Fisher information matrix \cite{paris2009}
\begin{equation}\label{10}
	F_{ij}=\frac{1}{2}\mathrm{Tr}\left\{\left(L_{\theta_i}L_{\theta_j}+L_{\theta_j}L_{\theta_i}\right)\rho\right\},
\end{equation}
where the symmetric logarithmic derivatives $\hat{L}_{\theta_i}$ satisfy the algebraic equations
\begin{equation}\label{11}
	2\partial_{\theta_i}\rho=\hat{L}_{\theta_i}\rho+\rho\hat{L}_{\theta_i}.
\end{equation}
Therefore, obtaining the QFIM from Eq.~\eqref{10} requires the determination of the symmetric logarithmic derivative $\hat{L}_{\theta_i}$ given in Eq.~\eqref{11}. Different explicit forms of the QFIM have been proposed in the literature \cite{banchi2014,sommers2003,paris2009}. If the density matrix is written through its spectral decomposition as
\[
\rho = \sum_k p_k |k\rangle \langle k|,
\]
then the QFIM can be expressed in terms of the eigenvalues and eigenvectors of $\rho$ as \cite{banchi2014,sommers2003}
\begin{equation} \label{12}
	F_{ij}=2\sum_{p_k+p_l>0}
	\frac{\langle k|\partial_{\theta_i}\rho|l\rangle
		\langle l|\partial_{\theta_j}\rho|k\rangle}
	{p_k+p_l},
\end{equation}
and the symmetric logarithmic derivatives take the form
\begin{equation}
	L_{\theta_i}=2\sum_{p_k+p_l>0}
	\frac{\langle k|\partial_{\theta_i}\rho|l\rangle}
	{p_k+p_l}
	|k\rangle\langle l|.
\end{equation}
The QFIM can also be formulated using the exponential representation of the density matrix \cite{paris2009}
\begin{equation} \label{14}
	F_{ij}=2\int_{0}^{\infty}
	\mathrm{Tr}\!\left[
	e^{-\rho t}\,(\partial_{\theta_i}\rho)\,
	e^{-\rho t}\,(\partial_{\theta_j}\rho)
	\right] dt .
\end{equation}
Recently, an explicit vectorized expression of the QFIM has been introduced in \cite{safranek2018}. This formulation is particularly convenient because it can be applied analytically to general quantum systems. In contrast to the spectral formula in Eq.~\eqref{12}, it does not require the diagonalization of the density matrix, and compared with Eq.~\eqref{14}, it avoids the calculation of matrix exponentials and integrals. The method is based on the inverse of the matrix
\begin{equation}\label{155}
	\Lambda = \left(\rho^{T}\otimes I + I\otimes \rho\right).
\end{equation}
Using the identities given in Eqs.~\eqref{6}, \eqref{7}, and \eqref{8}, the QFIM expressions reported in Eqs.~\eqref{12} and \eqref{14} can be recast as
\begin{equation}
	F_{ij}
	=
	2\,\mathrm{vec}\!\left[\partial_i \hat{\rho}\right]^{T}
	\Lambda^{-1}
	\mathrm{vec}\!\left[\partial_j \hat{\rho}\right].
\end{equation}
The symmetric logarithmic derivatives are then obtained from
\begin{equation} \label{17}
	\mathrm{vec}[L_{i}] = 2\Lambda^{-1}\mathrm{vec}[\partial_{\theta_i} \hat{\rho}].
\end{equation}
In the case of single-parameter estimation, the scalar quantum Cramér--Rao bound,
$\mathrm{Var}(\theta) \geq \mathcal{F}^{-1}$, can generally be reached. The corresponding optimal measurement is constructed from the projectors onto the eigenvectors of the symmetric logarithmic derivative $L_{\theta}$. However, the situation becomes more subtle in multiparameter estimation. Indeed, the matrix quantum Cramér--Rao bound,
\[
\mathrm{Cov}(\hat{\theta}) \geq F^{-1},
\]
is not always attainable. This is because the optimal measurements associated with different parameters may be incompatible with each other \cite{rehacek2018,ragy2016}. For this reason, it is necessary to identify the conditions under which the bound can be saturated.

To do so, Eq.~\eqref{17} must be solved in order to determine the symmetric logarithmic derivatives $L_{\theta_i}$ related to the different estimated parameters. If these operators commute, they share a common eigenbasis. In this case, a simultaneous measurement can be implemented, and the multiparameter Cramér--Rao bound can be saturated.

The condition
\[
[L_{\theta_i},L_{\theta_j}] = 0
\]
is sufficient, but it is not necessary. When the symmetric logarithmic derivatives do not commute, the weaker condition
\[
\mathrm{Tr}\left(\rho[L_{\theta_i},L_{\theta_j}]\right)=0
\]
is sufficient to ensure the attainability of the quantum Cramér--Rao bound \cite{ragy2016,matsumoto2002,crowley2014}.
\section{Physical Model}\label{sec:model}
We consider a monolayer graphene sheet lying in the $XOZ$ plane, such that the $y$-axis is perpendicular to the graphene layer. In this configuration, the electronic motion is confined to the $XOZ$ plane and the corresponding wave vector can be written as $\mathbf{k} = (k_x,0,k_z).$
Near the Dirac points $K$ and $K'$, the low--energy electronic excitations in graphene behave as massless Dirac fermions and can be described by an effective Dirac-like Hamiltonian.

For convenience and without loss of generality, we adopt natural units throughout
the following analysis by setting \(\hbar=v_F=k_B=1\). This choice simplifies the
analytical expressions and allows us to focus on the dependence of the physical
quantities on the relevant parameters of the system.

In order to keep the Hamiltonian consistent with the density matrix written below,
we use an effective four-dimensional representation of the low-energy graphene
model. The tensor-product notation used in this representation should not be
understood as describing two independent physical qubits. Rather, it provides a
convenient effective basis for the two-level degrees of freedom involved in the
model.

The first two-level degree of freedom is associated with the pseudospin structure
of graphene, which originates from the two sublattices \(A\) and \(B\) of the
honeycomb lattice. The second two-level degree of freedom accounts for the
twofold degeneracy of the effective low-energy description, which may be
associated with the valley sector or, equivalently, with a doubled representation
of the Dirac Hamiltonian. Therefore, the basis
\(\{|00\rangle,|01\rangle,|10\rangle,|11\rangle\}\) should be understood as an
effective computational basis used to write the Hamiltonian and the thermal
density matrix in a compact form.

Within this effective four-dimensional representation, the Hamiltonian is written as
\begin{equation}
	H =
	k_x\,\sigma_z\otimes I
	+
	k_z\,\sigma_x\otimes\sigma_x .
\end{equation}
where \(I\) is the \(2\times2\) identity matrix, while \(\sigma_x\) and
\(\sigma_z\) are the Pauli matrices defined by
\begin{equation}
	I=
	\begin{pmatrix}
		1 & 0 \\
		0 & 1
	\end{pmatrix},
	\qquad
	\sigma_x=
	\begin{pmatrix}
		0 & 1 \\
		1 & 0
	\end{pmatrix},
	\qquad
	\sigma_z=
	\begin{pmatrix}
		1 & 0 \\
		0 & -1
	\end{pmatrix}.
\end{equation}
Equivalently, in the basis
\(\{|00\rangle,|01\rangle,|10\rangle,|11\rangle\}\), it takes the matrix form
\begin{equation}
	H =
	\begin{pmatrix}
		k_z & 0 & k_x & 0 \\
		0 & -k_z & 0 & k_x \\
		k_x & 0 & -k_z & 0 \\
		0 & k_x & 0 & k_z
	\end{pmatrix}.
\end{equation}
Since \(H\) is centro-symmetric, it can be transformed into an X-form by means
of a double Hadamard transformation. We define
\begin{equation}
	U=H_{Had}\otimes H_{Had},
	\qquad
	H_{Had}=\frac{1}{\sqrt{2}}
	\begin{pmatrix}
		1 & 1 \\
		1 & -1
	\end{pmatrix}.
\end{equation}
The transformed Hamiltonian is then given by
\begin{equation}
	\mathcal{H}=UHU^\dagger .
\end{equation}
A direct calculation gives
\begin{equation}
	\mathcal{H}=
	\begin{pmatrix}
		k_x & 0 & 0 & k_z \\
		0 & k_x & k_z & 0 \\
		0 & k_z & -k_x & 0 \\
		k_z & 0 & 0 & -k_x
	\end{pmatrix}.
\end{equation}
The spectrum of this Hamiltonian is given by
\begin{equation}
	E_{\pm}=\pm\sqrt{k_x^2+k_z^2},
\end{equation}
with a twofold degeneracy. This degeneracy explains why the corresponding
thermal state can naturally be represented by a \(4\times4\) density matrix.

To investigate the thermodynamic properties of the system, we consider graphene
in thermal equilibrium at temperature \(T\). In this situation, the quantum state
of the system is described by the canonical density matrix
\begin{equation}
	\rho = \frac{e^{-\mathcal{H}/T}}{Z},
	\qquad
	Z=\mathrm{Tr}\left(e^{-\mathcal{H}/T}\right).
\end{equation}

In the standard computational basis $\{|00\rangle, |01\rangle, |10\rangle, |11\rangle\}$, the density matrix $\rho$ can be written in the following form
\begin{equation}
\rho =
\begin{bmatrix}
a & 0 & 0 & c \\
0 & a & c & 0 \\
0 & c & b & 0 \\
c & 0 & 0 & b
\end{bmatrix},
\end{equation}
where the matrix elements $a$, $b$ and $c$ depend on the physical parameters of the system such as the temperature $T$ and the components of the wave vector.
\begin{equation}
\begin{aligned}
a &= \frac{1}{4}\left(1-\frac{k_x \tanh\!\left(\frac{\sqrt{k_x^2+k_z^2}}{T}\right)}{\sqrt{k_x^2+k_z^2}}\right), \qquad
b = \frac{1}{4}\left(1+\frac{k_x \tanh\!\left(\frac{\sqrt{k_x^2+k_z^2}}{T}\right)}{\sqrt{k_x^2+k_z^2}}\right),\\[6pt]
c &= -\frac{k_z \tanh\!\left(\frac{\sqrt{k_x^2+k_z^2}}{T}\right)}{4\sqrt{k_x^2+k_z^2}} .
\end{aligned}
\end{equation}
In what follows, we consider the simultaneous estimation of the temperature \(T\)
and the wave--vector component \(k_x\), while \(k_z\) is kept fixed. To evaluate the quantum Fisher information matrix, one first needs to compute
the matrix $\Lambda$ defined in Eq.~\eqref{155}. It can be written as
\begin{equation}
\Lambda =
\begin{bmatrix}
\Lambda_{11} & \mathbf{0}_{4\times4} & \mathbf{0}_{4\times4} & \Lambda_{14} \\
\mathbf{0}_{4\times4} & \Lambda_{22} & \Lambda_{23} & \mathbf{0}_{4\times4} \\
\mathbf{0}_{4\times4} & \Lambda_{32} & \Lambda_{33} & \mathbf{0}_{4\times4} \\
\Lambda_{41} & \mathbf{0}_{4\times4} & \mathbf{0}_{4\times4} & \Lambda_{44}
\end{bmatrix}.
\end{equation}
with $\Lambda_{ij}$ $(i,j=1,2,3,4)$ the $4\times4$ matrices given by

\begin{equation}
\Lambda_{11}=\Lambda_{22}=
\begin{bmatrix}
2a & 0 & 0 & c \\
0 & 2a & c & 0 \\
0 & c & a+b & 0 \\
c & 0 & 0 & a+b
\end{bmatrix},
\qquad
\Lambda_{44}=\Lambda_{33}=
\begin{bmatrix}
a+b & 0 & 0 & c \\
0 & a+b & c & 0 \\
0 & c & 2b & 0 \\
c & 0 & 0 & 2b
\end{bmatrix},
\end{equation}

and

\begin{equation}
\Lambda_{23}=\Lambda_{32}=\Lambda_{41}=\Lambda_{14}=
\begin{bmatrix}
c & 0 & 0 & 0 \\
0 & c & 0 & 0 \\
0 & 0 & c & 0 \\
0 & 0 & 0 & c
\end{bmatrix}.
\end{equation}
The inverse of the matrix $\Lambda$ is given by
\begin{equation}
\Lambda^{-1} =
\begin{bmatrix}
(\Lambda^{-1})_{11} & \mathbf{0}_{4\times4} & \mathbf{0}_{4\times4} & (\Lambda^{-1})_{14} \\
\mathbf{0}_{4\times4} & (\Lambda^{-1})_{22} & (\Lambda^{-1})_{23} & \mathbf{0}_{4\times4} \\
\mathbf{0}_{4\times4} & (\Lambda^{-1})_{32} & (\Lambda^{-1})_{33} & \mathbf{0}_{4\times4} \\
(\Lambda^{-1})_{41} & \mathbf{0}_{4\times4} & \mathbf{0}_{4\times4} & (\Lambda^{-1})_{44}
\end{bmatrix}.
\end{equation}
with

with
\begin{align}
	(\Lambda^{-1})_{11}=(\Lambda^{-1})_{22}
	&=
	\begin{pmatrix}
		\alpha & 0 & 0 & \mu \\
		0 & \alpha & \mu & 0 \\
		0 & \mu & \delta & 0 \\
		\mu & 0 & 0 & \delta
	\end{pmatrix},
	&
	(\Lambda^{-1})_{33}=(\Lambda^{-1})_{44}
	&=
	\begin{pmatrix}
		\delta & 0 & 0 & \xi \\
		0 & \delta & \xi & 0 \\
		0 & \xi & \tau & 0 \\
		\xi & 0 & 0 & \tau
	\end{pmatrix},
\end{align}

and
\begin{align}
	(\Lambda^{-1})_{14}=(\Lambda^{-1})_{23}
	=(\Lambda^{-1})_{32}=(\Lambda^{-1})_{41}
	=
	\begin{pmatrix}
		\mu & 0 & 0 & \lambda \\
		0 & \mu & \lambda & 0 \\
		0 & \lambda & \xi & 0 \\
		\lambda & 0 & 0 & \xi
	\end{pmatrix}.
\end{align}

where the elements $\alpha$, $\delta$, $\xi$, $\mu$, $\lambda$ and $\tau$ are respectively given by
\begin{align}
	\alpha &= \frac{b(a+b)-c^{2}}{2(a+b)(ab-c^{2})},
	&
	\delta &= -\frac{-2ab+c^{2}}{2(a+b)(ab-c^{2})},
	&
	\xi &= \frac{ac}{2(a+b)(-ab+c^{2})},
	\\[0.4cm]
	\mu &= \frac{bc}{2(a+b)(-ab+c^{2})},
	&
	\lambda &= \frac{c^{2}}{2(a+b)(ab-c^{2})},
	&
	\tau &= \frac{a(a+b)-c^{2}}{2(a+b)(ab-c^{2})}.
\end{align}
Using the parameter derivatives of the density operator, the corresponding vectors can be written as
\begin{eqnarray}
	\mathrm{vec}\left[\partial_{k_x} \rho\right]
	&=&
	\left(\partial_{k_x} a,0,0,\partial_{k_x} c,0,\partial_{k_x} a,
	\partial_{k_x} c,0,0,\partial_{k_x} c,\partial_{k_x} b,0,\partial_{k_x} c,0,0,\partial_{k_x} b\right)^T,
	\\
	\mathrm{vec}\left[\partial_T \rho\right]
	&=&
		\left(\partial_{T} a,0,0,\partial_{T} c,0,\partial_{T} a,
	\partial_{T} c,0,0,\partial_{T} c,\partial_{T} b,0,\partial_{T} c,0,0,\partial_{T} b\right)^T.
\end{eqnarray}
The quantum Fisher information matrix associated with the two parameters
$\theta=(k_x,T)$ is then obtained as
\begin{eqnarray}
	\mathcal{F}
	=
	\left(
	\begin{array}{cc}
		\mathcal{F}_{k_x k_x} & \mathcal{F}_{k_x T} \\
		\mathcal{F}_{T k_x} & \mathcal{F}_{TT}
	\end{array}
	\right).
\end{eqnarray}

Explicitly, one finds
\begin{eqnarray}
	\mathcal{F}_{TT}
	&=&
\frac{
	\left(k_x^2+k_z^2\right)
	\mathrm{sech}^{2}\left(\frac{\sqrt{k_x^2+k_z^2}}{T}\right)
}{
	T^4
},
	\\
	\mathcal{F}_{k_x k_x}
	&=&
\frac{
	k_z^{2}T^{2}
	+
	\left(k_x^{4}+k_x^{2}k_z^{2}-k_z^{2}T^{2}\right)
	\mathrm{sech}^{2}\left(\frac{\sqrt{k_x^{2}+k_z^{2}}}{T}\right)
}{
	\left(k_x^{2}+k_z^{2}\right)^{2}T^{2}
},
	\\
	\mathcal{F}_{k_x T}
	&=&
	\mathcal{F}_{T k_x}
	=
-\frac{
	k_x\,\mathrm{sech}^{2}\left(\frac{\sqrt{k_x^{2}+k_z^{2}}}{T}\right)
}{
	T^{3}
}.
\end{eqnarray}

According to multiparameter quantum estimation theory, the covariance matrix
of any unbiased estimator is bounded from below by the inverse of the quantum
Fisher information matrix. Thus, for the parameter vector
$\theta=(k_x,T)$, the quantum Cramér--Rao inequality reads
\begin{eqnarray}
	\mathrm{Cov}(\theta)\geq \mathcal{F}^{-1}.
\end{eqnarray}

For a two-parameter estimation problem, the inverse matrix can be written as
\begin{eqnarray}
	\mathcal{F}^{-1}
	=
	\frac{1}{\det(\mathcal{F})}
	\left(
	\begin{array}{cc}
		\mathcal{F}_{TT} & -\mathcal{F}_{k_x T} \\
		-\mathcal{F}_{T k_x} & \mathcal{F}_{k_x k_x}
	\end{array}
	\right),
\end{eqnarray}
where
\begin{eqnarray}
	\det(\mathcal{F})
	=
	\mathcal{F}_{k_x k_x}\mathcal{F}_{TT}
	-
	\mathcal{F}_{k_x T}\mathcal{F}_{T k_x}.
\end{eqnarray}

Consequently, the lower bounds for the estimation variances are given by
\begin{eqnarray}
	\mathrm{Var}(k_x)
	&\geq&
	\frac{\mathcal{F}_{TT}}{\det(\mathcal{F})},
	\\
	\mathrm{Var}(T)
	&\geq&
	\frac{\mathcal{F}_{k_x k_x}}{\det(\mathcal{F})}\\
		\left(
		\operatorname{Var}(k_x)-\frac{\mathcal{F}_{TT}}{\det(\mathcal{F})}
		\right)
		\left(
		\operatorname{Var}(T)-\frac{\mathcal{F}_{k_xk_x}}{\det(\mathcal{F})}
		\right)
		&\geq&
		\left(
		\operatorname{Cov}(k_x,T)+\frac{\mathcal{F}_{k_xT}}{\det(\mathcal{F})}
		\right)^2 .
\end{eqnarray}
These expressions show that, in the simultaneous estimation scheme, the
precision bound of each parameter is affected not only by its own Fisher
information, but also by the correlations encoded in the off-diagonal elements
of the quantum Fisher information matrix. Therefore, the terms
$\mathcal{F}_{k_x T}$ and $\mathcal{F}_{T k_x}$ play an essential role in
controlling the achievable precision when $k_x$ and $T$ are estimated jointly. 
Using Eq.~\eqref{17}, we obtain the explicit matrix representations of the symmetric
logarithmic derivatives $L_{k_x}$ and $L_T$, which are given by
\begin{eqnarray}
	L_{k_x}=
	\left(
	\begin{array}{cccc}
		L_{k_x11} & 0 & 0 & L_{k_x33}\\
		0 & L_{k_x11} & L_{k_x33} & 0\\
		0 & L_{k_x33} & L_{k_x22} & 0\\
		L_{k_x33} & 0 & 0 & L_{k_x22}
	\end{array}
	\right),\quad 
		L_T=
	\left(
	\begin{array}{cccc}
		L_{T11} & 0 & 0 & L_{T33}\\
		0 & L_{T11} & L_{T33} & 0\\
		0 & L_{T33} & L_{T22} & 0\\
		L_{T33} & 0 & 0 & L_{T22}
	\end{array}
	\right).
\end{eqnarray}
where the matrix elements of the symmetric logarithmic derivatives $L_{k_x}$  and $L_T$ are defined as follows
\begin{eqnarray}
	L_{k_x11}
	&=&
	\frac{
		-k_x^2\sqrt{k_x^2+k_z^2}
		-\left(k_x^3+k_xk_z^2+k_z^2T\right)
		\tanh\left(\frac{\sqrt{k_x^2+k_z^2}}{T}\right)
	}
	{
		\left(k_x^2+k_z^2\right)^{3/2}T
	},
	\nonumber\\[2mm]
	L_{k_x22}
	&=&
	\frac{
		k_x^2\sqrt{k_x^2+k_z^2}
		-\left(k_x^3+k_xk_z^2-k_z^2T\right)
		\tanh\left(\frac{\sqrt{k_x^2+k_z^2}}{T}\right)
	}
	{
		\left(k_x^2+k_z^2\right)^{3/2}T
	},
	\nonumber\\[2mm]
	L_{k_x33}
	&=&
	k_xk_z
	\left[
	-\frac{1}{\left(k_x^2+k_z^2\right)T}
	+
	\frac{
		\tanh\left(\frac{\sqrt{k_x^2+k_z^2}}{T}\right)
	}
	{
		\left(k_x^2+k_z^2\right)^{3/2}
	}
	\right]\\
	L_{T11}
	&=&
	\frac{
		k_x+\sqrt{k_x^2+k_z^2}
		\tanh\left(\frac{\sqrt{k_x^2+k_z^2}}{T}\right)
	}
	{T^2},
	\nonumber\\[2mm]
	L_{T22}
	&=&
	\frac{
		-k_x+\sqrt{k_x^2+k_z^2}
		\tanh\left(\frac{\sqrt{k_x^2+k_z^2}}{T}\right)
	}
	{T^2},
	\nonumber\\[2mm]
	L_{T33}
	&=&
	\frac{k_z}{T^2}.\nonumber
\end{eqnarray}
From the analytical expressions of the SLD operators, one finds that the operators
associated with the temperature \(T\) and the wave vector component \(k_x\) do not
strictly commute. In particular, the commutator satisfies
\begin{equation}
	[L_T,L_{k_x}]\neq 0 .
\end{equation}
This result indicates that the optimal measurements associated with the two parameters
are not generally compatible at the operator level.

However, the strict commutation of the SLD operators is only a sufficient condition
for the saturability of the multiparameter quantum Cramér--Rao bound. In the present
case, the weaker compatibility condition is fulfilled,
\begin{equation}
	\mathrm{Tr}\left(\rho [L_T,L_{k_x}]\right)=0 .
\end{equation}
Therefore, although the SLD operators do not commute, the model remains compatible
in the weak sense. As a result, the simultaneous estimation scheme based on the
quantum Fisher information matrix remains meaningful, and the corresponding
multiparameter quantum Cramér--Rao bound can be attained.
\section{Precision Bounds in the Simultaneous Estimation Scheme} \label{sec:simultaneous}
This section is devoted to the numerical analysis of quantum coherence and multiparameter estimation precision in the considered system. We investigate how the temperature $T$ and the wave vector $k_x$ affect both the coherence $C$ and the minimum variances associated with parameter estimation. Particular emphasis is placed on the simultaneous estimation scheme, where correlations between the estimated parameters may significantly influence the achievable precision. 
\subsection{Temperature Dependence of Coherence and Estimation Precision}
In this subsection, we examine how temperature affects both the quantum coherence $C$ and the precision of temperature estimation. The aim is to determine whether the preservation of coherence is directly associated with an improvement in the estimation accuracy. For this purpose, we analyze the minimum variance obtained from the quantum Cramér--Rao bound for different values of the wave vector $k_x$. This allows us to identify the temperature regions where the system exhibits optimal metrological performance.\\
\begin{figure}[h!]
	\centering
	\includegraphics[width=1\textwidth]{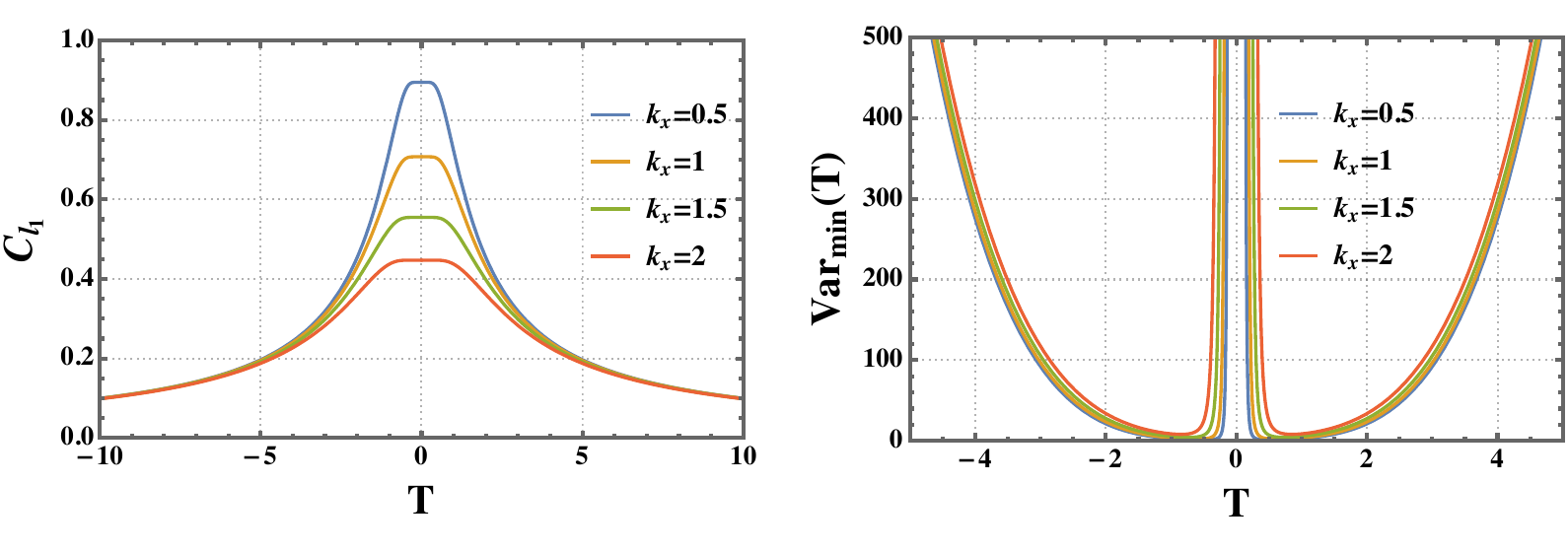}
\caption{Temperature dependence of quantum coherence and minimum estimation variance for different values of the wave vector $k_x$, with $k_z=1$.}
	\label{fig1}
\end{figure}
Figure~\ref{fig1} illustrates the behavior of the quantum coherence $C$ and the minimum variance $\mathrm{Var}^{\min}(T)$ as functions of the temperature $T$ for different values of the wave vector $k_x$. As shown in the left panel, the coherence reaches its maximum around $T=0$, indicating that the system preserves a strong quantum character in the low-temperature region. However, when $|T|$ increases, the coherence decreases progressively, which reflects the destructive effect of thermal fluctuations on quantum superposition.

Moreover, the maximum value of coherence decreases when $k_x$ increases. The highest coherence is obtained for $k_x=0.5$, whereas larger values of $k_x$ lead to a weaker coherence peak. This shows that the wave vector contributes to reducing the quantum coherence of the system.

The right panel displays the minimum variance $\mathrm{Var}^{\min}(T)$ associated with the estimation of temperature. A divergent behavior is observed around $T=0$, indicating that the temperature estimation becomes highly inaccurate in this region. This divergence means that, despite the high coherence near $T=0$, the system is not sufficiently sensitive to small variations of temperature, leading to a large Cramér--Rao bound.

Away from $T=0$, the variance decreases and reaches small values in intermediate temperature regions, where the estimation of $T$ becomes more precise. However, for large values of $|T|$, the variance increases again, showing that the estimation precision is also degraded at high temperatures. Therefore, the best temperature-estimation precision is obtained in an intermediate temperature range rather than at very low or very high temperatures.

In summary, this figure shows that the maximum coherence near $T=0$ does not necessarily correspond to the best temperature-estimation precision. Although the system exhibits strong coherence in this region, the variance $\mathrm{Var}^{\min}(T)$ diverges, indicating poor sensitivity to temperature variations. The optimal metrological performance is instead achieved at intermediate temperatures, where the variance is minimized. Thus, quantum coherence alone is not sufficient to guarantee precise temperature estimation; the sensitivity of the quantum state to the estimated parameter also plays a crucial role.
\subsection{Effect of the Wave Vector on Coherence and Estimation Precision}
We now analyze the influence of the wave vector $k_x$ on both the quantum coherence $C$ and the precision of wave-vector estimation. By comparing the behavior of $C$ with the minimum variance $\mathrm{Var}_{\mathrm{min}}^{}(k_x)$, we aim to determine whether strong coherence is associated with improved estimation accuracy. The analysis is performed for different values of the temperature $T$, allowing us to highlight the combined effects of thermal fluctuations and wave-vector variations on the metrological response of the system.

\begin{figure}[htbp]
	\centering
	\includegraphics[width=1\textwidth]{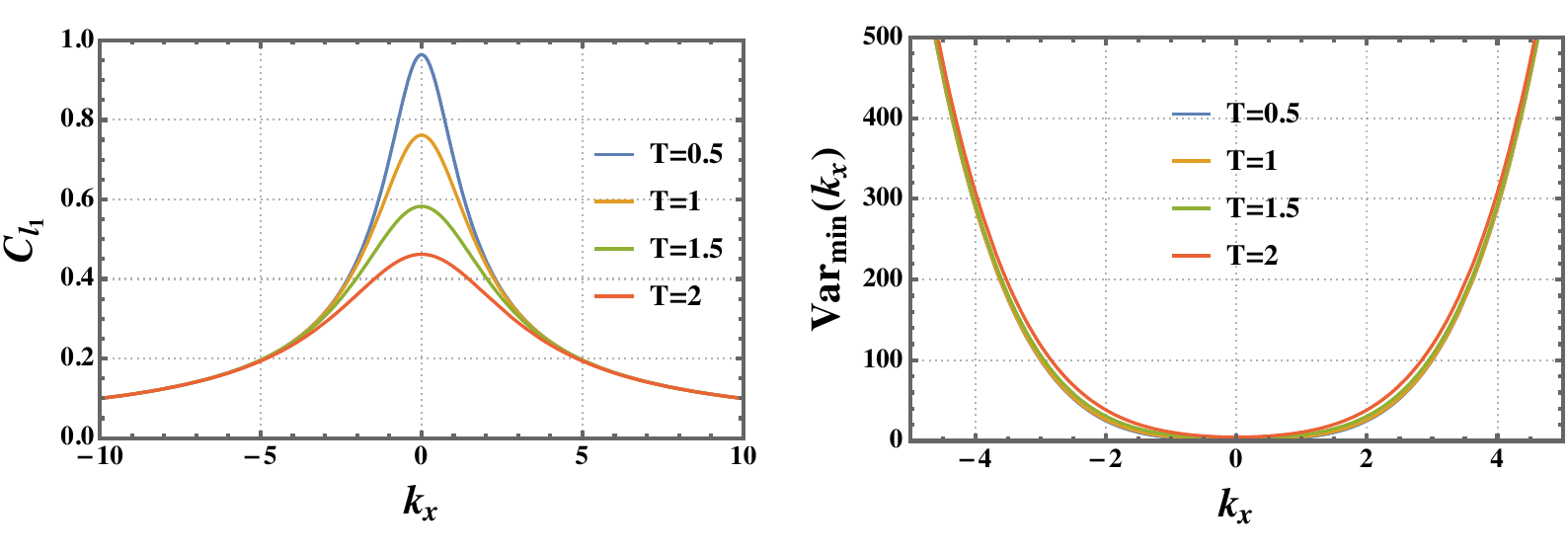}
\caption{Quantum coherence and minimum estimation variance in terms of the wave vector $k_x$ for different values of the temperature $T$, with $k_z=1$.}
	\label{fig2}
\end{figure}
In this second figure, the analysis is performed as a function of the wave vector $k_x$, for different values of the temperature $T$. In contrast to the previous figure, where the temperature was considered as the main variable, the present figure focuses on the direct effect of $k_x$ on quantum coherence and on the minimum estimation variances. This representation therefore provides a clearer view of how the wave-vector regime affects both the quantum behavior of the system and its metrological performance.

In the left panel, the coherence \(C\) is plotted as a function of \(k_x\) for
several fixed values of the temperature \(T\). The coherence reaches its maximum
around \(k_x=0\), which indicates that the system preserves a stronger quantum
character in the low-wave-vector region. As \(|k_x|\) increases, the coherence
decreases progressively, showing that large wave-vector values tend to weaken
the quantum superposition properties of the thermal state.

The effect of temperature is also clearly visible. Lower temperatures favor a
larger amount of coherence, whereas increasing \(T\) reduces the maximum value
of \(C\). This behavior reflects the destructive role of thermal fluctuations,
which tend to suppress the off-diagonal contributions of the density matrix.
Thus, quantum coherence is better preserved for low temperatures and small
values of \(k_x\).

In the right panel, the minimum variance \(\mathrm{Var}_{\min}(k_x)\) is shown as
a function of \(k_x\) for different values of \(T\). The variance takes its lowest
values around \(k_x=0\), indicating that this region is the most favorable for
estimating the wave-vector component with high precision. In contrast, when
\(|k_x|\) increases, the variance grows, which means that the estimation
precision becomes weaker.

Overall, this figure shows that the region around \(k_x=0\) is favorable for
both preserving quantum coherence and improving the precision of wave-vector
estimation. However, increasing the temperature reduces the coherence and
modifies the variance profile, confirming that thermal fluctuations play an
important role in the metrological response of the system.
\section{Precision Bounds in the Independent Estimation Scheme} \label{sec:independent}
Now, we turn to the case where the two parameters are estimated separately rather
than jointly. In this independent estimation scheme, each parameter is treated
through its own diagonal element of the quantum Fisher information matrix. If the
off-diagonal elements of the QFIM vanish, the parameters are fully decoupled and
the multiparameter bound reduces to independent scalar bounds. However, in the
present model, the off-diagonal term \(\mathcal{F}_{k_xT}\) is generally nonzero.
Therefore, the independent scheme considered here should be understood as a
separate-estimation reference strategy, in which the contribution of
\(\mathcal{F}_{k_xT}\) is not included in the single-parameter precision bounds.
This allows a direct comparison with the simultaneous estimation strategy, where
the full QFIM structure is taken into account.
\begin{eqnarray}
	\mathrm{Var}(k_x)^{\mathrm{Ind}} &\geq& F_{k_xk_x}^{-1}, \nonumber\\
	\mathrm{Var}(T)^{\mathrm{Ind}} &\geq& F_{TT}^{-1}.
\end{eqnarray}
When these two inequalities are saturated, one obtains
\begin{eqnarray}
	\mathrm{Var}(k_x)_{\min}^{\mathrm{Ind}}
	&=&
	\left(
	\frac{
		k_x^{2}T^{2}
		+
		\left(k_x^{4}+k_x^{2}k_z^{2}-k_z^{2}T^{2}\right)
		\sech\left(\frac{\sqrt{k_x^{2}+k_z^{2}}}{T}\right)
	}{
		\left(k_x^{2}+k_z^{2}\right)^{2}T^{2}
	}
	\right)^{-1},
	\\
	\mathrm{Var}(T)_{\min}^{\mathrm{Ind}}
	&=&
	\left(
	\frac{
		\left(k_x^2+k_z^2\right)
		\mathrm{sech}^2\left(\frac{\sqrt{k_x^2+k_z^2}}{T}\right)
	}{
		T^4
	}
	\right)^{-1}.
\end{eqnarray}
\subsection{Temperature Precision under Independent Estimation}
Before analyzing the independent estimation of temperature, it is useful to examine how quantum coherence behaves under the same conditions. This comparison allows us to clarify whether the regions where the system preserves a strong coherent character also correspond to better temperature sensitivity. In this context, the minimum variance associated with the independent estimation of $T$ is studied for different values of the wave vector $k_x$.

\begin{figure}[htbp]
	\centering
	\includegraphics[width=1\textwidth]{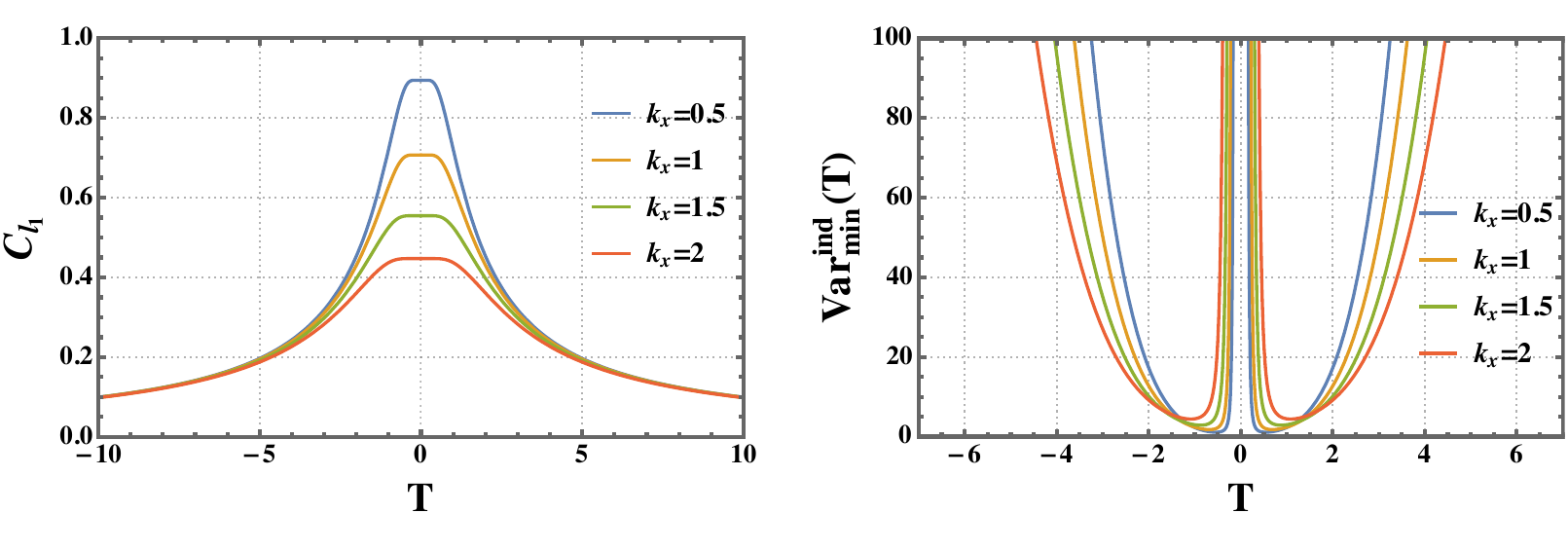}
	\caption{Temperature dependence of quantum coherence $C$ and the individual minimum estimation variance $\mathrm{Var}^{\mathrm{ind}}_{\min}(T)$ for different values of the wave vector $k_x$, with $k_z=1$.}
	\label{fig3}
\end{figure}
The figure \eqref{fig3} illustrates the influence of the temperature $T$ and the wave vector $k_x$ on the quantum coherence $C$ and the minimum variance of temperature estimation. As shown in the left panel, the coherence reaches its maximum near $T=0$, indicating that the system preserves stronger quantum superposition at low temperature. However, when $|T|$ increases, the coherence gradually decreases due to thermal effects. Moreover, increasing $k_x$ reduces the maximum value of coherence, showing that larger values of the wave vector weaken the quantum coherence of the system. 

The right panel shows that the minimum variance $\mathrm{Var}_{\mathrm{\min}}^{ind}(T)$ is highly sensitive to temperature. In particular, around $T=0$, $\mathrm{Var}_{\mathrm{\min}}^{ind}(T)$ exhibits a divergent behavior, which indicates a strong degradation of the temperature-estimation precision. This divergence is not related to the limitations of the simultaneous estimation strategy. 
Rather, it originates from the weak sensitivity of the thermal state to infinitesimal 
temperature variations in the low-temperature regime. As $T \to 0$, the thermal state becomes  weakly dependent on $T$, which leads to a small thermal Fisher information and consequently to a large Cramér--Rao bound. 

Away from this region, small values of $\mathrm{Var}_{\mathrm{\min}}^{ind}(T)$ indicate optimal temperature ranges where $T$ can be estimated with higher precision. The dependence on $k_x$ further demonstrates that the wave vector plays an important role in controlling the metrological performance of the system.

Overall, this figure reveals that quantum coherence and temperature-estimation precision do not necessarily follow the same behavior. Although the coherence is enhanced near $T = 0$, the independent estimation of temperature becomes less efficient in this region, as indicated by the divergence of $\mathrm{Var}^{\mathrm{ind}}_{\min}(T)$. This shows that a high degree of coherence does not automatically guarantee optimal metrological performance. Instead, the best estimation precision is achieved in intermediate temperature regions, where the system remains sufficiently sensitive to variations of $T$ while preserving an adequate quantum character. Therefore, in the independent estimation scheme, the relation between coherence and estimation precision is governed not only by the amount of coherence present in the system, but also by the sensitivity of the thermal state to temperature variations.
\subsection{Independent Wave-Vector Estimation and Quantum Coherence}
In this subsection, we focus on the independent estimation of the wave vector $k_x$ and its connection with quantum coherence. Since the parameters are treated separately in this scheme, the estimation precision is not affected by correlations with the temperature. This allows us to examine more clearly how the coherence $C$ influences the sensitivity of the system to variations of $k_x$ for different temperature values.

\begin{figure}[htbp]
	\centering
	\includegraphics[width=1\textwidth]{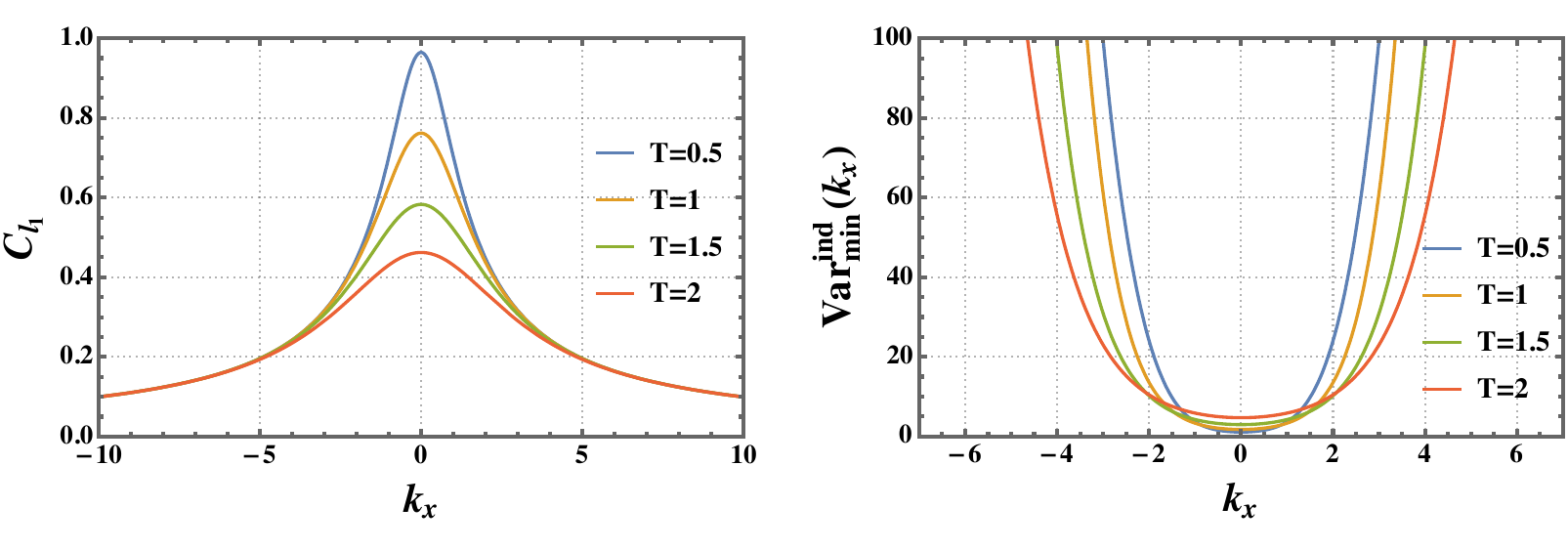}
	\caption{Wave-vector dependence of quantum coherence $C$ and the minimum variance $\mathrm{Var}_{\mathrm{ind}}^{\min}(k_x)$ in the independent estimation scheme for different values of the temperature $T$.}
	\label{fig4}
\end{figure}
Figure~\ref{fig4} illustrates the dependence of the quantum coherence $C$ and the minimum variance $\mathrm{Var}_{\mathrm{\min}}^{ind}(k_x)$ on the wave vector $k_x$ for different values of the temperature $T$. As shown in the left panel, the coherence reaches its maximum around $k_x=0$, indicating that the system preserves stronger quantum coherence in the low-wave-vector region. When $|k_x|$ increases, the coherence gradually decreases, showing that large values of the wave vector weaken the quantum superposition properties of the system. 

Moreover, increasing the temperature reduces the maximum value of coherence. The highest coherence is obtained for the lowest temperature, while higher temperatures lead to a progressive suppression of coherence. This behavior confirms that thermal effects tend to destroy the quantum character of the system.

The right panel shows the behavior of the minimum variance $\mathrm{Var}_{\mathrm{\min}}^{ind}(k_x)$ associated with the estimation of the wave vector $k_x$. The variance takes its smallest values around $k_x=0$, indicating that this region is the most favorable for estimating $k_x$ with high precision. In contrast, when $|k_x|$ increases, the variance grows rapidly, which means that the estimation precision becomes weaker. 

The dependence on temperature also shows that thermal effects modify the metrological performance of the system. In particular, increasing $T$ changes the width and the shape of the variance curves, indicating that the estimation of $k_x$ is strongly affected by temperature.

In summary, this figure shows that the optimal estimation of the wave vector $k_x$ is achieved around $k_x=0$, where the coherence reaches its maximum and the variance $\mathrm{Var}_{\mathrm{\min}}^{ind}(k_x)$ takes its minimum values. This indicates that, unlike the temperature-estimation case, quantum coherence appears to support the precision of $k_x$ estimation. However, increasing the temperature reduces the coherence and broadens the variance curves, leading to a degradation of the metrological performance. Therefore, the best compromise between coherence and estimation precision is obtained at low temperature and near $k_x=0$.\\
After analyzing the coherence and the estimation precision in both simultaneous and independent schemes, it is important to compare these two strategies more directly. Such a comparison allows us to evaluate the role of parameter correlations and to determine whether estimating the parameters jointly provides an advantage or, on the contrary, leads to a degradation of the overall precision. For this purpose, we introduce the ratio $\Gamma$, which quantifies the relative performance between the independent and simultaneous estimation schemes.
\section{Metrological Performance Ratio $\Gamma$} \label{sec:ratio}
To make the comparison between the two estimation strategies more transparent,
we introduce the dimensionless ratio \(\Gamma\). This quantity compares the total
estimation variance obtained in the simultaneous scheme with that obtained when
the parameters are treated independently. It therefore provides a useful indicator
of how the multiparameter structure affects the overall precision of the estimation
process. The ratio is defined as
\begin{equation}
	\Gamma =
	\frac{
		\mathrm{Var}_{\min}(k_x)
		+
		\mathrm{Var}_{\min}(T)
	}{
		2\left[
		\mathrm{Var}^{\mathrm{Ind}}_{\min}(k_x)
		+
		\mathrm{Var}^{\mathrm{Ind}}_{\min}(T)
		\right]
	}.
\end{equation}
After carrying out the algebraic simplification, we obtain
\begin{equation}
	\Gamma =
	\frac{1}{2}
	\left[
	1+
	\frac{
		k_x^{2}\left(k_x^{2}+k_z^{2}\right)
		\csch^{2}\left(\dfrac{\sqrt{k_x^{2}+k_z^{2}}}{T}\right)
	}{
		k_z^{2}T^{2}
	}
	\right].
\end{equation}
When \(\Gamma\) remains close to unity, the two schemes lead to comparable
estimation precision. In contrast, larger values of \(\Gamma\) indicate a more
pronounced difference between the two approaches, showing that the multiparameter
structure has a significant influence on the global metrological performance.\\
\begin{figure}[htbp]
	\centering
	\includegraphics[width=01\textwidth]{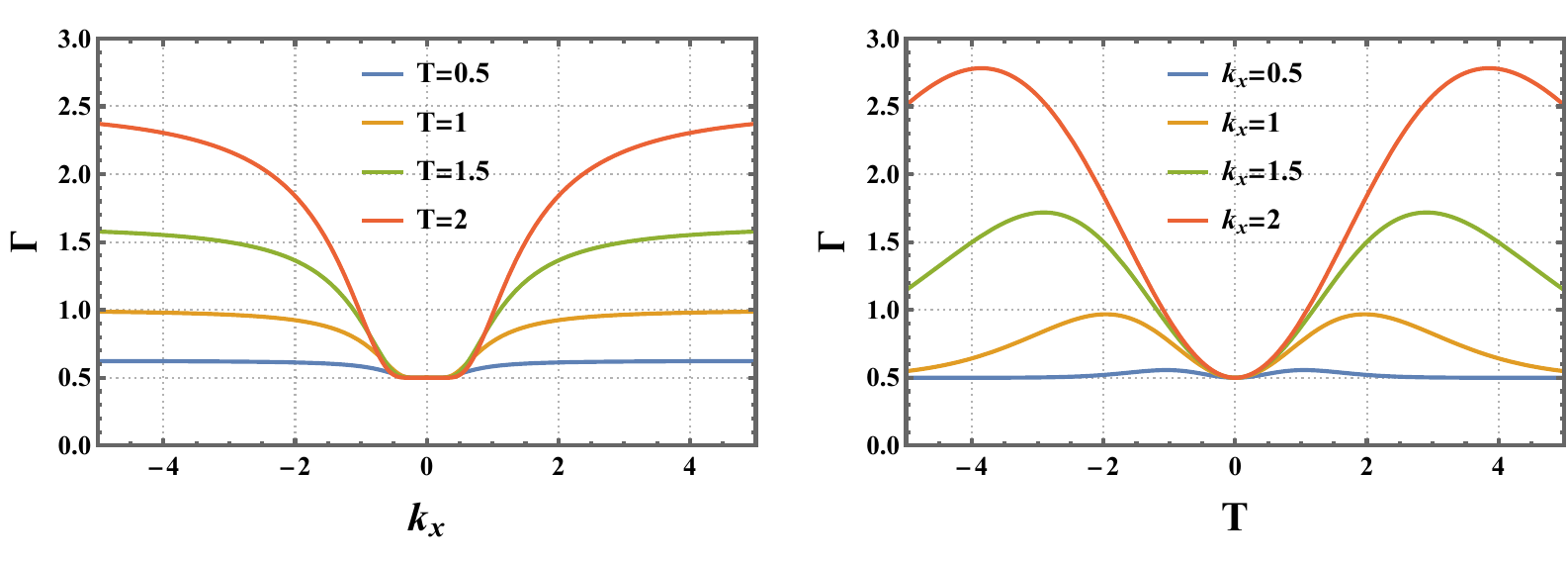}
\caption{Behavior of the metrological performance ratio $\Gamma$ as a function of the wave vector $k_x$ and the temperature $T$ in the comparison between the individual and simultaneous estimation schemes, with $k_z=1$.}
	\label{fig5}
\end{figure}
Figure~\ref{fig5} illustrates the evolution of the ratio \(\Gamma\) as a
function of the wave-vector component \(k_x\) and the temperature \(T\). It
shows how the multiparameter structure affects the relative performance of the
independent and simultaneous estimation schemes.

In the left panel, $\Gamma$ is represented as a function of $k_x$ for several fixed values of the temperature. It is clear that all curves exhibit a minimum around $k_x=0$, where the two estimation strategies become relatively close. This suggests that, in this region, the correlations between the parameters have only a limited effect on the estimation process. However, when $|k_x|$ increases, the ratio $\Gamma$ rises progressively, indicating that the difference between the independent and simultaneous approaches becomes more significant. This behavior becomes more pronounced at higher temperatures, showing that thermal fluctuations tend to amplify the influence of parameter correlations.

The right panel presents the variation of $\Gamma$ with respect to the temperature for different values of $k_x$. Around $T=0$, the ratio takes its lowest values, which confirms that the two estimation schemes behave almost similarly in the low-temperature region. As $|T|$ increases, $\Gamma$ increases as well, especially for larger values of $k_x$. This means that the combined effect of temperature and wave vector enhances the contribution of parameter correlations to the global estimation precision. The highest values of $\Gamma$ are obtained for the largest wave vectors, confirming that $k_x$ plays an important role in strengthening the difference between the two strategies.

Overall, the results show that the ratio $\Gamma$ is strongly affected by both $T$ and $k_x$. Around $T=0$ and $k_x=0$, $\Gamma$ reaches its minimum value, indicating that the difference between the independent and simultaneous estimation schemes is less pronounced in this region. In contrast, increasing either the temperature or the wave vector leads to larger values of $\Gamma$, suggesting that the simultaneous estimation strategy becomes more advantageous when the two parameters are treated jointly rather than separately.
\section{Conclusion and Perspectives} \label{sec:conclusion}
In this work, we have studied the interplay between quantum coherence and multiparameter quantum estimation in a graphene-based system. The analysis focused on the estimation of the temperature $T$ and the wave vector component $k_x$ by using the quantum Fisher information matrix and the quantum Cramér--Rao bound. Both simultaneous and independent estimation schemes were considered in order to clarify the role of parameter correlations in the achievable precision.

Our results show that quantum coherence is mainly enhanced in the low-temperature regime and around $k_x=0$. However, increasing the temperature or the absolute value of the wave vector progressively reduces the coherence, reflecting the destructive influence of thermal fluctuations and wave-vector effects on quantum superposition. Nevertheless, the regions of maximum coherence do not always correspond to the best estimation precision. In particular, the variance associated with temperature estimation diverges near $T=0$, showing that the system becomes weakly sensitive to small temperature variations in this region. Therefore, the optimal estimation of $T$ is obtained at intermediate temperatures rather than in the regime of maximum coherence.

In contrast, the estimation of the wave vector $k_x$ is more directly connected to the coherence properties of the system. The minimum variance of $k_x$ reaches its lowest values around $k_x=0$, where the coherence is also maximal. This indicates that quantum coherence can support wave-vector estimation, especially at low temperatures. However, thermal effects still reduce this advantage by weakening coherence and increasing the estimation variance.

The comparison between the simultaneous and independent estimation schemes further shows that parameter correlations play an important role in the global metrological performance. This effect was quantified through the ratio $\Gamma$, which reveals that the two strategies behave similarly near $T=0$ and $k_x=0$, while their difference becomes more pronounced for larger temperatures and wave-vector values. Thus, the multiparameter nature of the problem must be taken into account when evaluating the precision limits.

Overall, our findings demonstrate that quantum coherence is a useful resource for enhancing metrological precision, but it is not sufficient on its own to guarantee optimal estimation. The performance also depends on the sensitivity of the quantum state to the estimated parameter and on the correlations between parameters. As a perspective, this work could be extended by including additional effects such as magnetic fields, spin--orbit coupling, disorder, or decoherence mechanisms, in order to explore more realistic graphene-based quantum sensing platforms.


\begin{thebibliography}{99}
\bibitem{novoselov2004}
K. S. Novoselov, A. K. Geim, S. V. Morozov, D. Jiang, Y. Zhang, S. V. Dubonos, I. V. Grigorieva, and A. A. Firsov, "Electric Field Effect in Atomically Thin Carbon Films," \textit{Science}, \textbf{306}, 666 (2004).

\bibitem{geim2007}
A. K. Geim and K. S. Novoselov, "The rise of graphene," \textit{Nature Materials}, \textbf{6}, 183 (2007).

\bibitem{castro2009}
A. H. Castro Neto, F. Guinea, N. M. R. Peres, K. S. Novoselov, and A. K. Geim, "The electronic properties of graphene," \textit{Reviews of Modern Physics}, \textbf{81}, 109 (2009).

\bibitem{helstrom1976}
C. W. Helstrom, \textit{Quantum Detection and Estimation Theory}, Academic Press, New York (1976).

\bibitem{paris2009}
M. G. A. Paris, "Quantum Estimation for Quantum Technology," \textit{International Journal of Quantum Information}, \textbf{7}, 125 (2009).

\bibitem{baumgratz2014}
T. Baumgratz, M. Cramer, and M. B. Plenio, "Quantifying Coherence," \textit{Physical Review Letters}, \textbf{113}, 140401 (2014).

\bibitem{ragy2016}
S. Ragy, M. Jarzyna, and R. Demkowicz-Dobrzański, "Compatibility in multiparameter quantum metrology," \textit{Physical Review A}, \textbf{94}, 052108 (2016).

\bibitem{liu2020}
J. Liu, H. Yuan, X.-M. Lu, and X. Wang, "Quantum Fisher information matrix and multiparameter estimation," \textit{Journal of Physics A: Mathematical and Theoretical}, \textbf{53}, 023001 (2020).

\bibitem{gilchrist2009}
A. Gilchrist, D. R. Terno, and C. J. Wood, "Vectorization of quantum operations and its use," \textit{arXiv preprint}, arXiv:0911.2539 (2009).

\bibitem{schacke2004}
K. Schacke, "On the Kronecker product," Master's thesis, University of Waterloo, 2004.


\bibitem{banchi2014}
L. Banchi, P. Giorda, and P. Zanardi, "Quantum information-geometry of dissipative quantum phase transitions," \textit{Physical Review E}, \textbf{89}, 022102 (2014).

\bibitem{sommers2003}
H. J. Sommers and K. Życzkowski, "Bures volume of the set of mixed quantum states," \textit{Journal of Physics A: Mathematical and General}, \textbf{36}, 10083 (2003).

\bibitem{safranek2018}
D. \v{S}afr\'anek, "Simple expression for the quantum Fisher information matrix," \textit{Physical Review A}, \textbf{97}, 042322 (2018).

\bibitem{rehacek2018}
J. \v{R}eh\'a\v{c}ek, Z. Hradil, D. Koutn\'y, J. Grover, A. Krzic, and L. L. S\'anchez-Soto, "Optimal measurements for quantum spatial superresolution," \textit{Physical Review A}, \textbf{98}, 012103 (2018).


\bibitem{matsumoto2002}
K. Matsumoto, "A new approach to the Cram\'er--Rao-type bound of the pure-state model," \textit{Journal of Physics A: Mathematical and General}, \textbf{35}, 3111 (2002).

\bibitem{crowley2014}
P. J. Crowley, A. Datta, M. Barbieri, and I. A. Walmsley, "Tradeoff in simultaneous quantum-limited phase and loss estimation in interferometry," \textit{Physical Review A}, \textbf{89}, 023845 (2014).


\end{thebibliography}
\end{document}